\documentclass[aps,preprint, floatfix]{revtex4}

\usepackage{graphicx,color}
\usepackage{psfrag}
\usepackage{amssymb}
\usepackage{eso-pic}
\usepackage[dvips,letterpaper]{geometry} 
\usepackage{amsmath}

\begin{document}

%
%
%
%
\def\oti{{\otimes}}
\def\lb{ \left[ }
\def\rb{ \right]  }
\def\tilde{\widetilde}
\def\bar{\overline}
\def\hat{\widehat}
\def\*{\star}
\def\[{\left[}
\def\]{\right]}
\def\({\left(}		\def\BL{\Bigr(}
\def\){\right)}		\def\BR{\Bigr)}
	\def\BBL{\lb}
	\def\BBR{\rb}
%
%
\def\zb{{\bar{z} }}
\def\zbar{{\bar{z} }}
\def\frac#1#2{{#1 \over #2}}
\def\inv#1{{1 \over #1}}
\def\half{{1 \over 2}}
\def\d{\partial}
\def\der#1{{\partial \over \partial #1}}
\def\dd#1#2{{\partial #1 \over \partial #2}}
\def\vev#1{\langle #1 \rangle}
\def\ket#1{ | #1 \rangle}
\def\rvac{\hbox{$\vert 0\rangle$}}
\def\lvac{\hbox{$\langle 0 \vert $}}
\def\2pi{\hbox{$2\pi i$}}
\def\e#1{{\rm e}^{^{\textstyle #1}}}
\def\grad#1{\,\nabla\!_{{#1}}\,}
\def\dsl{\raise.15ex\hbox{/}\kern-.57em\partial}
\def\Dsl{\,\raise.15ex\hbox{/}\mkern-.13.5mu D}
%
%
\def\ga{\gamma}		\def\Ga{\Gamma}
\def\be{\beta}
\def\al{\alpha}
\def\ep{\epsilon}
\def\vep{\varepsilon}
\def\la{\lambda}	\def\La{\Lambda}
\def\de{\delta}		\def\De{\Delta}
\def\om{\omega}		\def\Om{\Omega}
\def\sig{\sigma}	\def\Sig{\Sigma}
\def\vphi{\varphi}

%
%
\def\CA{{\cal A}}	\def\CB{{\cal B}}	\def\CC{{\cal C}}
\def\CD{{\cal D}}	\def\CE{{\cal E}}	\def\CF{{\cal F}}
\def\CG{{\cal G}}	\def\CH{{\cal H}}	\def\CI{{\cal J}}
\def\CJ{{\cal J}}	\def\CK{{\cal K}}	\def\CL{{\cal L}}
\def\CM{{\cal M}}	\def\CN{{\cal N}}	\def\CO{{\cal O}}
\def\CP{{\cal P}}	\def\CQ{{\cal Q}}	\def\CR{{\cal R}}
\def\CS{{\cal S}}	\def\CT{{\cal T}}	\def\CU{{\cal U}}
\def\CV{{\cal V}}	\def\CW{{\cal W}}	\def\CX{{\cal X}}
\def\CY{{\cal Y}}	\def\CZ{{\cal Z}}

\def\rvac{\hbox{$\vert 0\rangle$}}
\def\lvac{\hbox{$\langle 0 \vert $}}
\def\comm#1#2{ \BBL\ #1\ ,\ #2 \BBR }
\def\2pi{\hbox{$2\pi i$}}
\def\e#1{{\rm e}^{^{\textstyle #1}}}
\def\grad#1{\,\nabla\!_{{#1}}\,}
\def\dsl{\raise.15ex\hbox{/}\kern-.57em\partial}
\def\Dsl{\,\raise.15ex\hbox{/}\mkern-.13.5mu D}
%
%
%
\font\numbers=cmss12
\font\upright=cmu10 scaled\magstep1
\def\stroke{\vrule height8pt width0.4pt depth-0.1pt}
\def\topfleck{\vrule height8pt width0.5pt depth-5.9pt}
\def\botfleck{\vrule height2pt width0.5pt depth0.1pt}
\def\Zmath{\vcenter{\hbox{\numbers\rlap{\rlap{Z}\kern
0.8pt\topfleck}\kern 2.2pt
                   \rlap Z\kern 6pt\botfleck\kern 1pt}}}
\def\Qmath{\vcenter{\hbox{\upright\rlap{\rlap{Q}\kern
                   3.8pt\stroke}\phantom{Q}}}}
\def\Nmath{\vcenter{\hbox{\upright\rlap{I}\kern 1.7pt N}}}
\def\Cmath{\vcenter{\hbox{\upright\rlap{\rlap{C}\kern
                   3.8pt\stroke}\phantom{C}}}}
\def\Rmath{\vcenter{\hbox{\upright\rlap{I}\kern 1.7pt R}}}
\def\Z{\ifmmode\Zmath\else$\Zmath$\fi}
\def\Q{\ifmmode\Qmath\else$\Qmath$\fi}
\def\N{\ifmmode\Nmath\else$\Nmath$\fi}
\def\C{\ifmmode\Cmath\else$\Cmath$\fi}
\def\R{\ifmmode\Rmath\else$\Rmath$\fi}

\def\barray{\begin{eqnarray}}
\def\earray{\end{eqnarray}}
\def\beq{\begin{equation}}
\def\eeq{\end{equation}}

\def\sc{\scriptstyle}

\def\n{\noindent}

\def\kvec{{\bf{k}}}
\def\gradvec{\vec{\nabla}}
\def\Evec{\vec{E}}
\def\Bvec{\vec{B}}
\def\vplus{\oplus}
\def\vminus{\ominus}
\def\vphi{\varphi}
\def\vphibar{\bar{\varphi}}
\def\smallhalf{{\scriptstyle \inv{2}}}
\def\phi{\Phi}
\def\Etilde{\tilde{E}}   
\def\xitilde{\tilde{\xi}}
\def\ftilde{\tilde{f}}

\def\up{\uparrow}
\def\down{\downarrow}
\def\smallhalf{{\textstyle \inv{2}}}
\def\smallsqrt{{\textstyle \inv{\sqrt{2}}}}
\def\AA{\leavevmode\setbox0=\hbox{h}\dimen0=\ht0 \advance\dimen0 by-1ex\rlap{
\raise.67\dimen0\hbox{\char'27}}A}

\title{An electrostatic depiction  of the validity of the  
 Riemann Hypothesis  
 and a formula for the N-th zero at large N}
\author{ Andr\'e  LeClair}
\affiliation{Cornell University,  Physics Department, Ithaca, NY 14850}

\bigskip\bigskip\bigskip\bigskip

\begin{abstract}

We construct a vector   field $\Evec$ from the real and imaginary parts 
of an entire   function $\xi (z)$  which arises  in the quantum statistical mechanics of relativistic gases when 
the spatial dimension $d$ is analytically continued into the complex $z$ plane.  This function is 
built from the $\Gamma$ and Riemann $\zeta$ functions and is known to satisfy the  functional identity
$\xi (z) = \xi (1-z)$.      $\Evec$ satisfies the conditions for a static electric field.   
The structure of   $\Evec$ in the critical strip is determined by its behavior near 
the Riemann zeros on the  critical line $\Re (z) = 1/2$,  where each zero can be assigned a 
$\vplus $ or $\vminus$  parity, or vorticity,   of a related  pseudo-magnetic field.   Using these  properties,     we show that  
a  hypothetical Riemann zero that is off the critical line  leads to a frustration of this ``electric'' field. 
   We formulate this frustration  more precisely  in terms of  the potential $\phi$ satisfying $\Evec = - \gradvec \phi$ and construct $\phi$ explicitly.    
The main outcome  of our analysis is a formula for the $n$-th zero  on the critical line for large $n$ 
expressed as the solution of a simple transcendental equation.   Riemann's counting formula for the number
of zeros on the entire critical strip can be derived from this formula.   
Our  result is much  stronger  than Riemann's  counting formula,  since it provides an estimate of the $n$-th zero
along the critical line.   
 This  provides a simple way to estimate very high zeros to very good accuracy,  and we estimate the
$10^{10^6}$-th one.

\end{abstract}

\maketitle

\section{Introduction}   

 Riemann's  zeta function\cite{Edwards}   was originally defined as the infinite series\cite{sigma}. 
\beq
\label{ZetaDef}
\zeta(z)  \equiv  \sum_{n=1}^\infty    \inv{n^z}    ~~~~~~~\Re(z) > 1  
\eeq
It can be analytically continued throughout the complex $z$ plane.   
    We will refer to roots $\rho$   of the equation $\zeta (\rho)  =0$  as ``Riemann zeros'',  or simply as ``zeros''.    
There are  trivial zeros   for  $z$ equal to any  negative even integer.  
It has been proven that there are no zeros along the line $\Re (z) =1$,    which is equivalent to proving the 
Prime Number Theorem,  as proven by Hadamard and  and de  la Vall\'ee  Poussin.    
It is also known that there are an  infinite number of zeros along the ``critical line''  $\Re(z) = \smallhalf$,  which was  proven by Hardy.         
The Riemann Hypothesis (RH) is the statement that the latter  are the  only zeros within the ``critical strip'' 
$0 \leq \Re (z) \leq 1$.    
Riemann's major result  was an explicit formula,  expressed in terms of these zeros,   which describes the distribution of prime numbers.   However the RH,  and other several other statements in the original paper,  are still  
unresolved\cite{Edwards}.     We refer the reader to  Conrey's article  for a short, but 
excellent  introduction to the RH itself\cite{Conrey}.

An important aspect of this subject concerns the counting of zeros.   Riemann estimated that the 
 number $N$  of zeros on the strip, i.e. with  $0\leq  \Re (\rho) \leq 1$,   and $0\leq \Im (\rho) \leq T$,  as 
\beq
\label{Nzeros}
N(T)   \approx    \frac{T}{2\pi} \log \( \frac{T}{2\pi} \)  - \frac{T}{2\pi}  
\eeq
There are known corrections to the above formula,  involving $\arg \, \zeta$,  which we  will rederive  below.  
Riemann did not provide a proof,  but it was eventually proven  by von Mangoldt about 45  years later.  
It has never  been proven that the above formula applies to the counting of zeros on the critical line.    
The reason being that  this  would essentially   prove   the RH,  or at the very least   certainly would follow from the RH.  
We will present a proof that $N(T)$ applies to the zeros on the critical line,  and more,  by deriving a simple  formula for the $N$-th  zero on the 
critical line  for  large $N$,  eqs. (\ref{transcends}, \ref{FinalTranscendence}) below.

There have been a number of elaborate attempts to prove the RH using physical ideas,  but unfortunately  without  success as far as an actual proof of the RH.   (One was proposed by us.)     Many of these approaches attempt to explain the explicit values of the Riemann zeros, 
based for instance  on the Hilbert-P\'olya  idea that the zeros are the spectrum of some as yet  unknown hamiltonian,  in particular the work of Berry and Keating,   who suggested that
the hamiltonian must have chaotic behavior,  and some extensions due to Sierra\cite{Berry1, BerryKeating, BerryKeating2, Sierra,  Sierra2}.     A related but essentially different approach is
due to Connes\cite{Connes}. 
   For an extensive review,  we refer to \cite{Schumayer} and references therein.    Some more 
   recent work not reviewed there is by Sierra and collaborators\cite{Sierra1,Sierra3}.    
   These  ideas  are very interesting,   and may perhaps  eventually lead to a proof of the RH. 
        In any case,  we were led  to suspect  that the resolution lies in traditional real and complex analysis rather than physics or  arithmetic.    Based on reading his original paper,   Riemann appeared  to be confident of this point of view\cite{Rpaper}.

It  may be argued that the area of physics where Riemann's zeta function plays the most prominent and direct  role is
in the  quantum statistical mechanics of gases.     Although we will not be invoking ideas from statistical physics to  
study  the details of the  RH here,  it is nevertheless instructive to use this connection as a way of describing some of the important properties of
$\zeta(z)$.   It is this connection  that initiated our interest in the problem in the first place,   however we should state
from the outset that we  will not 
bring any  methods based intrinsically on physics  to bear on the problem.

Consider a gas of massless,  relativistic bosons with single-particle energy $E_\kvec = |\kvec|$ in $d$ spatial 
dimensions,   where $\kvec$ is  the momentum vector,   at a  temperature $T= 1/\beta$ and zero chemical potential.    
The free energy density,  which is minus the pressure,  is given by the
well-known  formula which can be found in any elementary textbook on quantum statistical physics:
\beq
\label{FreeEnergy}
\CF = \inv{\beta}  \int  \frac{d^d \kvec}{(2\pi)^d}  ~ \log \( 1- e^{-\beta |\kvec|} \)  = -\inv{\beta^{d+1}} \frac{\Gamma(d) \zeta(d+1) } {2^{d-1} \pi^{d/2} \Gamma (d/2) }   ,  ~~~~~\Re (d) > 0 
\eeq
Here $\Gamma$ is the standard Euler gamma function satisfying $\Gamma (z+1) = z \Gamma (z)$.   
When the boson in question is one polarization of a  physical  photon,  then the above formula leads to Planck's black body spectrum,
and this  discovery  in fact marked the birth of  Quantum Mechanics;  Planck's constant 
 $\hbar$ was first determined in this way.     
In performing the above integral we used 
$\int d^d \kvec =  \frac{2 \pi^{d/2}} {\Gamma(d/2)} \int d k   \, k^{d-1}  $,  where $k= |\kvec|$.   
In many other physical  situations,  in order to regularize divergent integrals,  one analytically continues certain functions of 
the dimension $d$ into the complex plane in a  procedure referred to as ``dimensional regularization'',  and we will do the same here in this Introduction.

In the path integral approach to quantum field theory at finite temperature,  the above free energy corresponds to 
the logarithm of a functional integral over a scalar  field which is a function of  $d+1$ dimensional spatial coordinates 
 after analytically  continuing to euclidean time  $t\to - i \tau$ and compactifying $\tau$ so that it lives on a circle of
 circumference $\beta$.     (For $d=1$ this would be an infinitely long cylinder.)    
 The same path integral can be viewed as a zero temperature quantum mechanical problem where 
 the compact $\tau$ direction is now regarded  as a spatial coordinate and time is an infinite line,   i.e. euclidean time and one spatial coordinate
 are interchanged in comparison to the finite temperature picture. 
     The path integral defining $\CF$ now corresponds to a Casimir energy,  i.e. 
 the ground state energy density with one compactified spatial coordinate.    This leads to a very different  
 divergent expression
 which must be regulated using the $\zeta$ function:  
\beq
\label{Casimir}
\CE_0  =  \inv{2\beta}  \sum_{n\in \Zmath}   \int  \frac{ d^{d-1} \kvec}{(2\pi)^{d-1}} ~ \sqrt{  \kvec^2 + (2 \pi n \beta)^2 }
=   - \inv{\beta^{d+1}} \pi^{d/2} \Gamma(-d/2) \zeta(-d) 
\eeq
Physically  these two expressions must be equal,  which is to say,   $\CF = \CE_0$,  and this  leads to 
the primary functional identity satisfied by the $\zeta$ function,  based on a modular
 transformation\cite{AL}.    Namely,    define 
$\chi (z) = \pi^{-z/2} \Gamma(z/2)  \zeta(z) $.       Using the doubling formula for the 
$\Gamma$ function   $\Gamma(d)  = 2^{d-1} \Gamma(d/2) \Gamma((d+1)/2) / \sqrt{\pi}  $,   one has 
$- \beta^{d+1}  \CF =  \chi (d+1) $ and 
$ - \beta^{d+1}  \CE_0 = \chi (-d) $.     Upon analytically continuing $d$,   these two expressions are equal due to the 
functional identity   $\chi(z) = \chi (1-z) $   The latter  identity is well-known and there
are many different purely mathematical proofs of it;   
in fact it  was known to Riemann.

The fact that the above functional identity can  be obtained  by this very different quantum mechanical argument  indicates  its   non-triviality.      This leads one to think that this identity plays a central role in 
 establishing the validity of the RH.   For instance,  
the trivial zeros at $z=-2n$ follow simply from this functional identity since it implies
$\zeta (-2n) = \pi^{-2n-1/2}  \Gamma(n+\smallhalf)  \zeta (1 + 2n)/ \Gamma(-n)$  and $\Gamma(z)$ has a pole at
$z=-n$.   
 From this functional relation one can also obtain results such as $\zeta(4) = \pi^4/90$, relevant for  black body physics,       and similar expressions involving the Bernoulli numbers   for all
$z$ equal to a positive even integer.  Thus many of the most  important properties of $\zeta$ follow simply  from  the functional identity for $\chi$.       Furthermore,   this identity is satisfied by
other  Dirichlet  L-functions which are believed to also satisfy a RH,  which is one aspect  of 
Langland's program.

The function $\zeta (z) $ has only one pole,  a simple pole at $z=1$.     Incidentally,   this pole is the reason why Bose-Einstein condensation is well  known to be 
 impossible in two dimensions,  based on the work of  Coleman,   Mermin and Wagner\cite{Coleman,Mermin}.      The density of non-relativistic bosons with $E_\kvec = \kvec^2 /2m$ is
\bigskip
\beq
\label{BEC}
n =  \int  \frac{ d^d \kvec}{(2\pi)^d}  \inv{  e^{\beta \kvec^2 /2m} -1 }  = 
\( \frac{m}{2 \pi \beta} \)^{d/2}   \zeta (d/2) 
\eeq  
and diverges when $d=2$.    
 
It is thus convenient to multiply $\chi (z)$  by $z(z-1)$ to remove this pole.   We therefore define,  as Riemann did,   the function
\beq
\label{xidef}
\xi (z) \equiv   \smallhalf z (z-1) \chi (z) =   \smallhalf  z (z-1)   \pi^{-z/2}  \Gamma( z/2)  \zeta (z)
\eeq
     The function $\xi (z)$ has the same zeros as $\zeta (z)$,  and 
also satisfies  the important   identity
\beq
\label{identity}
 \xi (z) = \xi (1-z)
 \eeq  
 $\xi (z)$ is an entire function,   which is to say it is single valued,  analytic,  and differentiable everywhere
in the complex plane.    

Having made these introductory remarks,   let us now summarize the main work  of this article.   
The $\zeta$ function is difficult to visualize because it is a map from the complex plane into 
itself,  and  requires a four dimensional plot to display all of its properties.
    In order to visualize the function, 
we construct a vector field $\Evec$ from the real and imaginary parts of $\xi (z)$.       By virtue of
the Cauchy-Riemann equations this field satisfies the conditions for a static electric field with no charged  sources, namely
it has zero divergence and curl.    We emphasize that our analysis is not based on physical principles but
only  on  mathematics;  
the analogy with electrostatics is simply a useful one,  and guided our investigation.     We will thus refer to $\Evec$ as the
``electric field''.   We  apologize to mathematician readers who may find this a distraction, but they
may readily ignore the terminology;  
we included this electrostatic visualization since it was instrumental to our initial understanding of the problem
and may perhaps lead to further developments.

The vector field 
  $\Evec$  can be expressed as the gradient of an electric potential $\phi$,   $\Evec = - \gradvec \phi$.
Since $\phi$ is a real function,    this provides a more economical way to visualize the RH.    
This is described in  section III,   where it is shown that the RH would follow  from some
simple properties of $\phi$ along the line $\Re (z) =1$,  which we refer to as  a ``regular alternating''  property of a real
function.       

In order to establish this regular alternating property requires some analysis of the detailed properties of $\xi (z)$,
which are presented in section  IV.  
      A by-product of this analysis is a characterization of the 
zeros that is stronger than Riemann's counting estimate,  namely,  that the n-th zero is of the form
$\rho_n = \smallhalf  + i y_n$    where $y_n$ satisfies the simple transcendental equation, 
 (\ref{FinalTranscendence}) or approximately 
  by the explicit formula eq. (\ref{Lambert}) in terms of the Lambert function $W$.   
     This is the main analytic result of our work.        This new formula works extremely well,  as Table \ref{Table1} below shows.
If desired,  the reader may actually jump directly to section IV   where this equation satisfied
by the $n$-th zero is derived in a self-contained way.

We wish to stress from the beginning that  although equations (\ref{FinalTranscendence}) 
and (\ref{counting2}) have an obvious resemblance,   they are fundamentally different
in meaning.    Eq. (\ref{FinalTranscendence}) was derived {\it on the critical line},  whereas 
(\ref{counting2}) is a staircase function that has only been proven for the entire {\it critical strip},
and furthermore by  assuming that $T$ is the not the ordinate of a zero.   
As we will show,  one  can derive (\ref{counting2}) from (\ref{FinalTranscendence}),  but not visa versa.    
The result    (\ref{FinalTranscendence})  represents an infinite number of \emph{equations},  
  which  depend only on $n$,  and they determine   the imaginary
parts $y_n$'s of the Riemann zeros.      In other words, 
the $n$-th Riemann zero is the unique solution of these equations.

\section{Electrostatic  analogy  and visualization of the RH.}

Let us define the real and imaginary parts of $\xi(z)$ as 
\beq
\label{realimag}
\xi (z) =  u(x,y) + i \, v(x,y)
\eeq
where $z= x + i y$.   
 The Cauchy-Riemann equations,  $\d_x u = \d_y v$ and $\d_y u = - \d_x v$, 
are satisfied everywhere since $\xi$ is an entire function.  
   Consequently,   both $u$ and $v$ are harmonic functions,   i.e.  solutions of the Laplace equation 
$\gradvec^2 u  =  (\d_x^2 + \d_y^2 ) u = 0$,     $\gradvec^2 v = 0$,  although they are not completely independent.    
Let us define $u$  or $v$  contours as the curves in the $x,y$ plane corresponding to $u$ or $v$ equal to
a constant,  respectively. 
The critical line is a $v=0$ contour since $\xi$ is real along it.     
As a consequence of the Cauchy-Riemann equations,    $\gradvec u \cdot \gradvec v =0$.
Thus,  where the $u,v$ contours intersect,  they are  necessarily perpendicular,  and this is one aspect
of their dependency.     A Riemann zero occurs wherever the  $u=0$ and $v=0$ contours intersect.  

\begin{figure}[htb] 
\begin{center}
\hspace{-15mm} 
\psfrag{xh}{$\scriptstyle x=1/2 $}
\psfrag{xo}{$\scriptstyle x=0 $}
\psfrag{x1}{$\scriptstyle x=1 $}
\psfrag{x}{$\sc x $}
\psfrag{y}{$\sc y$}
\psfrag{vo}{$\scriptstyle v=0$}
\psfrag{uo}{$\scriptstyle u={\rm constant}$}
\psfrag{vg}{$\scriptstyle v>0$}
\psfrag{vl}{$\scriptstyle v<0$}
\includegraphics[width=6cm]{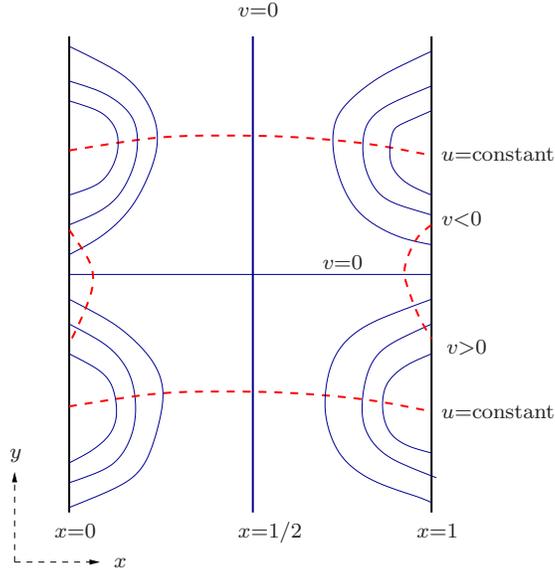} 
\end{center}
\caption{Constant $u$ contours are dashed lines (red on-line)  and $v$ contours are solid lines (blue on-line).
Riemann zeros occur where a $u=0$ contour spans the entire strip and crosses the critical line.}
\vspace{-2mm}
\label{uvcontours} 
\end{figure}

     From  the symmetry eq. (\ref{identity})
 and $\xi (z)^* = \xi (z^*)$  it follows that 
\beq
\label{symmetries}
 u(x,y) = u (1-x, y), ~~~~~ 
v(x,y) = - v (1-x,y)
\eeq   
  This implies that the $v$ contours do not cross the critical line except for $v=0$.    
All  the $u$ contours on the other hand are allowed to cross it by the above symmetry.    
Away from the $v=0$ points on the line $\Re (z) = 1$,   since the $u$ and $v$ contours are perpendicular,  
   the $u$ contours generally cross the critical line and span the whole strip due to the symmetry eq. (\ref{symmetries}).     The $u$ contours that do not cross the critical line must be in the vicinity of the $v=0$ contours,
   again by the perpendicularity of their intersections.         Putting all  these facts together,  
the $u,v$ contours on the critical strip have the properties shown in Figure \ref{uvcontours}.  
This demonstrates that Riemann zeros indeed exist on the critical line.     Hardy proved that there
are infinitely many of such zeros,  and we will refer to  this fact later.

Introduce the vector field 
\beq
\label{Evec}
\vec{E} =   E_x  \, \hat{x}  +  E_y  \, \hat{y}  \equiv   u (x,y) \,  \hat{x}   -   v(x,y) \,   \hat{y}
\eeq
where $\hat{x}$ and $\hat{y}$ are unit vectors in the $x$ and $y$ directions.     
One purpose of this article is to map out the properties of this field and describe  their implications  
for  the RH.       

This field has zero divergence and curl as a consequence of the Cauchy-Riemann equations
\beq
\label{curl}
\gradvec \cdot \Evec    =0 , ~~~~~\gradvec \times \Evec = 0, 
\eeq
which are defined everywhere since $\xi$ is entire.   
Thus it satisfies the conditions of a static electric field with no charged sources.   We will continue to make 
this analogy and refer to $\Evec$ as the ``electric field''.    However we
  wish to emphasize that 
 although such  analogies  will be invoked in the sequel,    $\Evec$ is not 
a physically realized electric field here,  in that  we do not need to specify what kind of charge distribution 
would give rise to such a field.     All of our  subsequent arguments will be based only on the mathematical identities expressed in 
eq. (\ref{curl}),  and our reference to electrostatics is simply a useful analogy,  as stated in the Introduction.      
Since the divergence of $\Evec$ equals zero everywhere,  the hypothetical electric charge distribution that gives
rise to $\Evec$ should be thought of as existing at infinity.    Alternatively,  since $u$ and $v$ are harmonic
functions,   one can view them as being determined by their values on the boundary of the critical strip.     
 Although less meaningful,  it is useful  to also consider   a  ``magnetic field'' 
$\vec{B}  =    u (x,y) \,  \hat{x}   +    v(x,y) \,   \hat{y}$.     Here $\gradvec \times \vec{B} \neq 0$,  so it
is analogous to a magnetic field with non-zero currents as sources.   However it is not a proper magnetic field
since $\gradvec \cdot \vec{B} \neq 0$;  nevertheless it will be useful merely  for establishing some terminology.

As we now argue,  the main properties of the above $\Evec$  field on the critical strip are determined by its behavior near the 
Riemann zeros on the critical line combined with the behavior near $\Re(z) =1$.        The $\Evec$  field can be expressed as a gradient of an electric potential
$\Evec = -\gradvec \phi$,  and satisfies the usual properties of a physical electric field with no sources. 
In particular,  electric field lines do not cross. 
We will reconstruct $\phi$ below,  which leads to a more precise argument.      

\begin{figure}[htb] 
\begin{center}
\hspace{-15mm} 
\psfrag{x}{$\sc  x=1/2 $}
\psfrag{vo}{$\sc v=0$}
\psfrag{uo}{$\sc u=0$}
\psfrag{E}{$\vec{E}$}
\psfrag{B}{$\vec{B}$}
\includegraphics[width=8cm]{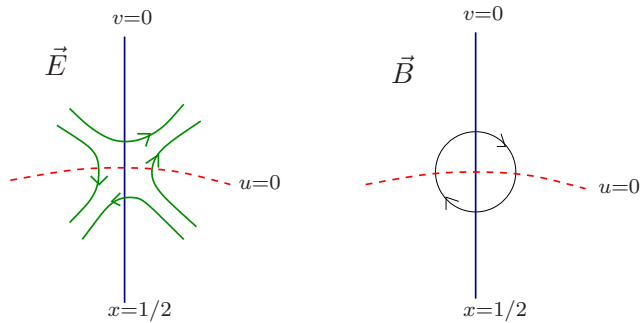} 
\end{center}
\caption{Sketches of the ``electric''    and ``magnetic''  fields $\vec{E}$, $\vec{B}$,  in the vicinity of a Riemann zero 
 along the critical line identified as a $\vplus$ vortex.   A zero of type $\vminus$  has the direction of all arrows reversed.}
\vspace{-2mm}
\label{vortex1} 
\end{figure}

\begin{figure}[htb] 
\begin{center}
\hspace{-15mm} 
\includegraphics[width=6cm]{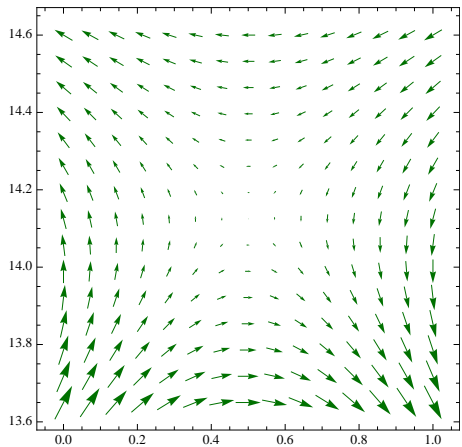} 
\end{center}
\caption{The   field $\Evec$ in the vicinity of the first  Riemann zero $z= 1/2 + 14.1347\,  i   $.   
This zero has $\vminus$ vorticity.}
\vspace{-2mm}
\label{vectorplot} 
\end{figure}

\begin{figure}[htb] 
\begin{center}
\hspace{-15mm} 
\includegraphics[width=6cm]{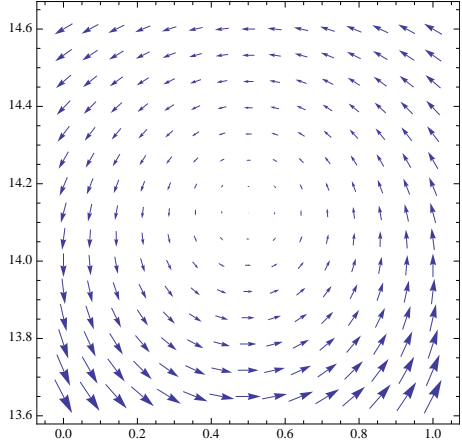} 
\end{center}
\caption{The field $\Bvec$ in the vicinity of the Riemann zero $z= 1/2 + 14.1347\,  i   $.}
\vspace{-2mm}
\label{Bvectorplot} 
\end{figure}

  Any Riemann zero on the critical line arises from a $u=0$ contour
that crosses the full width of the strip and thus intersects the vertical $v=0$ contour.  
   On the $u=0$ contour,  $E_x=0$,  whereas on the $v=0$ contour of
the critical line itself,  $E_y =0$.     Furthermore,  $E_y$ changes direction as one crosses the critical line.  
Finally,  taking  into account that $\Evec$ has zero curl,   one can easily see that there are only two ways that 
all these conditions can be satisfied near the Riemann zero.      
Simply using the fact that the $y$ component of $\Bvec$ has  the opposite sign of $\Evec$,  
one can see that the $\vec{B}$ field encircles the Riemann zero in either the clockwise or anti-clockwise direction.
Thus each Riemann zero can be assigned a vorticity, or parity,    $\vplus $ or $\vminus$   where $\vplus$ refers to
clockwise.     These properties are sketched in Figure \ref{vortex1}.    
In short,  Riemann zeros on the critical strip are  manifestly consistent with the necessary properties of $\Evec$.     
The  vector plots   of $\Evec$ and $ \Bvec$  for the actual function $\xi (z)$  in the vicinity of
the first Riemann zero at $z= 1/2 +  14.1347 \, i$   are shown in 
Figures \ref{vectorplot}, \ref{Bvectorplot},  and    
confirm  these simple arguments. 
 
In  this picturesque analogy,    the Riemann zeros along the critical line 
behave like  very small regions with nearly constant potential $\phi$.  
Since physically  $\phi$ is constant inside a conductor,   the Riemann zeros are analogous to very  thin  conducting wires  that penetrate the strip perpendicularly  to it,  and  are  electrically neutral.     
The wires can also  be thought to 
carry a current that  gives  rise to $\gradvec \times \Bvec \neq 0$,  however as stated earlier,  the analogy with
a bonafide physical magnetic field is not  correct,   since $\gradvec \cdot \Bvec \neq 0$,  and 
$\gradvec \times \Bvec$ is not singular at the Riemann zeros.      In other words,  there is no analog of
Ampere's law for $\Bvec$.   

\begin{figure}[htb] 
\begin{center}
\hspace{-15mm} 
\psfrag{xh}{$\sc x=1/2 $}
\psfrag{xo}{$\sc x=0 $}
\psfrag{x1}{$\sc x=1 $}
\psfrag{x}{$\sc x $}
\psfrag{y}{$\sc y$}
\psfrag{vo}{$\sc v=0$}
\psfrag{uo}{$\sc u=0$}
\psfrag{z1}{$\sc \rho_1$}
\psfrag{z2}{$\sc \rho_2$}
\includegraphics[width=6cm]{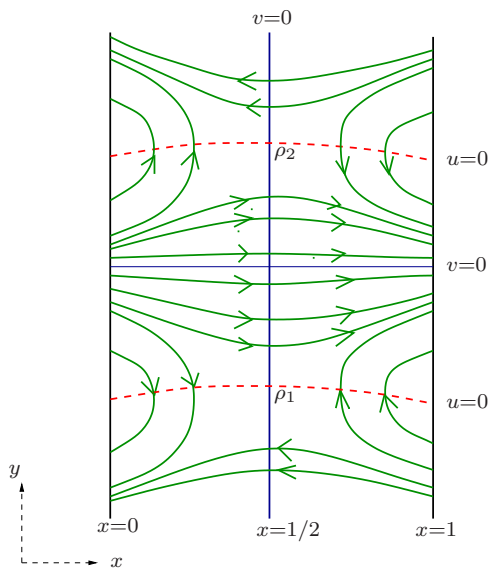} 
\end{center}
\caption{The field $\vec{E}$ (in green on-line) in the vicinity of two consecutive Riemann zeros $\rho_1 , 
\rho_2$  on the
critical line.}
\vspace{-2mm}
\label{Electric} 
\end{figure}

We now turn to the global properties of $\Evec$ along the entire critical strip.     
It is not difficult to see  that the vorticity of the Riemann zeros on the critical line, $\vplus$ or $\vminus$,  
alternate in sign as one moves along it.    Otherwise,   the curl of $\Evec$ would not be zero 
in a  region between two consecutive zeros.   
 Thus there is a form of quasi-periodicity along the 
critical line,  in the sense that zeros alternate between being even and odd,  like the integers,  and also  analogous to
the zeros of $\sin (x)$ at $x= \pi n$  where $e^{i \pi n} = (-1)^n$.
A proof of this alternating vorticity requires some analysis of the function $\xi$,  which is
described in section IV.    
   Also,  along  the nearly horizontal $v=0$ contours that cross 
the critical line, 
$\Evec$ is in the $x$ direction.    This leads to the pattern in Figure \ref{Electric}.   
One aspect of  the rendition of this pattern is that it implicitly assumes that the $v=0$ and $u=0$ points along the line
$\Re (z) =1$ alternate,  namely,  between two consecutive $v=0$ points along this line,  there is only one
$u=0$ point,  which is consistent with the knowledge that there are no zeros of $\xi$ along the line $\Re (z) =1$.   
This fact will be clearer when we reformulate our argument in terms of the potential $\phi$ below.

\begin{figure}[htb] 
\begin{center}
\hspace{-15mm} 
\psfrag{xh}{$\sc x=1/2 $}
\psfrag{x1}{$\sc x=1 $}
\psfrag{uo}{$\sc \sc u=0$}
\psfrag{vo}{$\sc v=0$}
\includegraphics[width=3cm]{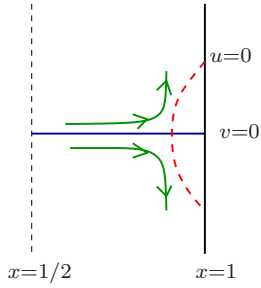} 
\end{center}
\caption{ The frustrated electric field $\vec{E}$ (in green on-line)  in the vicinity of a hypothetical Riemann zero off the critical line.}
\vspace{-2mm}
\label{frustrated} 
\end{figure}

Let us now  consider the possibility of a Riemann zero elsewhere on the critical strip, i.e. off  of the critical line. 
The only location this could occur is along a $v=0$ contour that intersects the critical line.    In order for a Riemann
zero to exist there requires a $u=0$ contour to intersect the $v=0$ contour,  and recall they must be perpendicular
at the intersection.   Such a $u$ contour,  of the kind that does not cross the critical line,   is shown in 
Figure \ref{uvcontours}.     In comparison with the zeros on the critical line,  the $u$ and $v$ contours are nearly  rotated by
$90^\circ$.    This leads to a ``frustration'' of the electric field.    Namely,   as Figure \ref{Electric} indicates,  
$\Evec$ wants to be horizontal at such a location,   but if there exists a $u=0$ contour here,  this would imply
the electric field would be vertical along it,  as shown in Figure \ref{frustrated}.          Since the pattern in Figure 
\ref{Electric}  very likely repeats itself all along the critical strip because of the known infinity of zeros along it,     this frustration appears to be  inconsistent,   and this suggests  that  such a Riemann zero would appear to be  impossible,  although
this is not yet a proof.      One needs to establish that the pattern in Figure \ref{Electric}  indeed regularly repeats itself
all along the critical strip.  

   The RH is equivalent to
the statement that all $u=0$ contours cross the critical line and span  the entire strip.   We have argued that  this  follows from the existence of the infinite number of zeros along the critical line,   the identity eq. (\ref{identity}),   and the global properties of the field $\Evec$ which are a consequence  of  the existence of these zeros.     
To strengthen this picture,  one needs to more accurately define the notion of  ``frustration'' in this context.
In the next section,  a more concrete  analysis will be based on the potential $\phi$.   
A related  remark is  that the properties of $\Evec$ depicted in Figure \ref{Electric}  require  that the $v=0$ and $u=0$ points
along the line $\Re (z) = 1$ are well separated,  and this is consistent with  the proven fact that there are no
roots to the equation $\zeta (1 + i y) =0$,   which can be used  to prove the Prime Number Theorem\cite{Edwards}.

\section{The electric potential $\phi$}

A mathematically integrated   version of the above arguments,  which 
has the advantage  of making manifest the dependency  of $u$ and $v$, 
 can be formulated in terms of the  electric potential $\phi$ which is a single real function.      
  Although it contains the same information as the above argument,  it is more economical,   and 
  more importantly,  because it    does not rely on properly defining the notion 
of ``frustration''  in order to make the argument.           
By virtue of $\gradvec \cdot \Evec =0$,  $\phi$ is also a solution of Laplace's equation 
$\d_z \d_\zbar \phi =0$ where $\zbar =  z^*$.    The general solution is  
that $\phi$ is the sum of a function of $z$ and another function of $\zbar$.    Since $\phi$ must be real, 
\beq
\label{pot}
\Evec =   - \gradvec \phi,  ~~~~~  \phi (x,y) =  \smallhalf  \(   \varphi (z)   + \vphibar (\zbar) \) 
\eeq
where $\vphibar (\zbar) =  \varphi (z)^*$.    Clearly $\phi$ is not analytic,  whereas 
$\varphi$ is;    it is useful to work with $\phi$ since we only have to deal with one real function.  
 Comparing the definitions of $\Evec$ and $\xi$ in terms of $u,v$,  one 
finds $u = -(\d_z \vphi + \d_\zbar \vphibar )/2$ and $v= - i ( \d_\zbar \vphibar - \d_z \vphi)/2$.    The latter implies 
\beq
\label{Ed}   
\xi (z) = - \frac{\d \varphi (z)}{\d z  }
\eeq
This equation can be integrated because $\xi $ is entire.   
Using an  integral representation for $\xi (z)$ derived in Riemann's original paper, one can show  that  up to 
an  irrelevant  additive constant,  
\beq
\label{phi}
\varphi (z) = - 8 \int_1^\infty  d[ t^{3/2} g'(t) ]  ~   \frac{t^{-1/4}}{\log t}   \,    \sinh \[ \smallhalf (z-\smallhalf) \log t \]  \eeq
where   $g'(t)$ is the $t$-derivative of the function $g(t) =  \smallhalf \( \vartheta_3 (0, e^{-\pi t}) -1\)  =  \sum_{n=1}^\infty e^{-n^2 \pi t}$,  and  $\vartheta_3$ is one of the four elliptic theta functions.

Let us now consider the $\phi = {\rm constant}$ contours in the critical strip.   Using the integral 
representation eq. (\ref{phi}),  one finds the symmetry 
 $\phi (x,y) = -\phi(1-x, y)$.   One sees then  that the $\phi \neq 0$ contours 
do not cross the critical line,   whereas the $\phi =0$ contours can and do.    Since $\vphi$ is imaginary along 
the critical line,   the latter is also a $\phi=0$ contour.    Thus the $\phi$ contours have the same
structure as those for $v$ shown in Figure \ref{uvcontours}.   

All  Riemann zeros  $\rho$ necessarily occur at isolated points,  which is a property of entire functions.
This is clear from the factorization formula $\xi (z) = \xi (0)  \prod_\rho   ( 1- z/\rho )$,  conjectured by Riemann, 
and later proved  by Hadamard.    Where are these zeros located in terms of $\phi$?   
At $\rho$,  $\gradvec \phi =0$.     Thus, such isolated zeros occur when two $\phi$ contours 
{\it intersect},   which 
can only occur if the two contours correspond to the same value of $\phi$ since $\phi$ is single-valued.    A useful analogy is the electric potential for equal point charges.   The electric field vanishes halfway between them,  and this is the unique point where the equi-potential contours vanish.   
The argument is simple:     $\gradvec \phi$ is perpendicular to the $\phi$ contours,  however as one approaches
$\rho$ along one contour,  one sees that it is not in the same direction as inferred from the approach from the other contour.  The only way this could be consistent is if $\gradvec \phi = 0$ at $\rho$.     

\begin{figure}[htb] 
\begin{center}
\hspace{-15mm} 
\includegraphics[width=6cm]{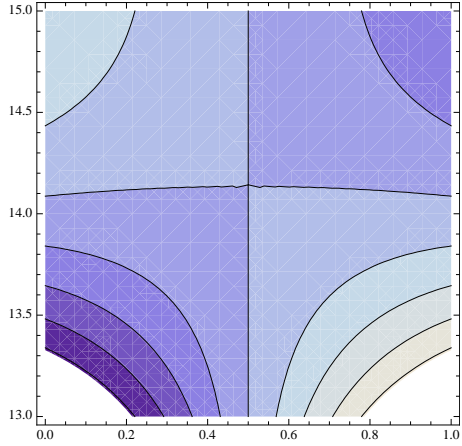} 
\end{center}
\caption{ Contour plot of the potential $\phi$  in the vicinity of the first Riemann zero  at $z=1/2 + 14.1347 \, i$.  
The horizontal,  vertical directions are the $x,y$ directions where $z=x + i y$.   
The critical line and nearly horizontal line are $\phi=0$ contours and they intersect at the zero.}
\vspace{-2mm}
\label{phicontour} 
\end{figure}

With these properties of $\phi$,  we can now begin to  understand the location of the known  Riemann zeros.   
Since the $\phi=0$ contours intersect  the critical line,  which is also a $\phi=0$ contour,  a zero 
exists at each such intersection,  and we know there are an infinite number of them.   
The contour plot in Figure \ref{phicontour} for the actual function $\phi$ constructed above verify these statements.   We emphasize that there is nothing special about the value $\phi =0$, 
since $\phi$ can be shifted by an arbitrary constant without changing $\Evec$;  we defined it
such that the critical line corresponds to $\phi =0$.    

\begin{figure}[htb] 
\begin{center}
\hspace{-15mm} 
\psfrag{po}{$\phi = 0$}
\psfrag{pno}{$\phi \neq 0$}
\psfrag{rn1}{$\rho_{n+1}$}
\psfrag{rn}{$\rho_n$}
\psfrag{xh}{$x=\smallhalf$}
\psfrag{x1}{$x=1$}
\includegraphics[width=6cm]{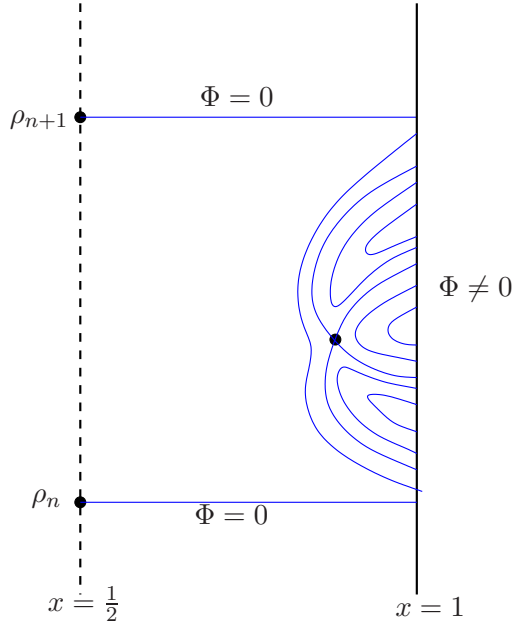} 
\end{center}
\caption{ A sketch of the contour plot of the potential $\phi$  in the vicinity of  a hypothetical Riemann zero off of the critical line.
Such a zero occurs where the contours intersect.  $\rho_n$ and $\rho_{n+1}$ are consecutive zeros on the 
line.  }
\vspace{-2mm}
\label{phicross} 
\end{figure}

A  hypothetical  Riemann zero off of the critical line would then necessarily  correspond to an intersection of two 
$\phi \neq 0$ contours.   For simplicity,  let us assume that only two such contours intersect,  since our arguments can be
easily extended to more of such intersections.     Such a situation is depicted in Figure \ref{phicross}. 
This figure   implies that on the line $\Re( z) =1$,  specifically  $z=1+iy$,    $\phi$  takes on  the same non-zero value at four different values  of $y$ between consecutive  zeros,  i.e. roots  of the equation $ f(y) = 0$,
where $f(y) \equiv \phi(1, y) = 
\Re (\vphi (1+iy))$.   Thus,  the  real function
$f(y)$ would have to have 3 extrema between two consecutive zeros.     Figure \ref{phicontour} suggests
that this does not occur.   In order to begin to prove it,  let us define a ``regular alternating"  real function  $h(y)$  of a real variable 
$y$ as a function that alternates between positive and negative values  in the most regular manner possible:  between
two consecutive zeros $h(y)$ has only one maximum, or minimum.     For example,  the  $\sin (y)$ function is obviously  regular  alternating.    By the above argument,   if $f(y)$  is regular alternating,
then two $\phi \neq 0$ contours cannot intersect and there are no Riemann zeros off the critical line.

To summarize this section,   based on the symmetry eq. (\ref{identity}),  and the existence of the known infinity 
of Riemann zeros along the critical line,   we have argued that $\Evec$ and $\phi$ satisfy a regular repeating
pattern all along the critical strip,  and the RH would follow from such a repeating pattern.       In order to go further,   one needs to investigate the  detailed properties of the function
$\xi$,  in particular it's large $y$ asymptotic behavior,  and establish its repetitive behavior, 
more specifically,  that it is a regular alternating function. 
This is the subject of the next section,   which is
more constructive.

\section{Analysis and an asymptotic formula for the $N$-th Riemann zero.}

\def\smallelev{{\scriptstyle  \frac{11}{8}}}
\def\chitilde{\hat{\chi}}

As in the Introduction,  define $\chi  (z) = \pi^{-z/2} \Gamma (z/2) \zeta (z)$,   which also 
has the symmetry $\chi(z) = \chi (1-z)$.     
It has a pole at $z=1$,  but this does not affect the analysis of the  zeros for large $\Im (z)$ 
on the critical strip. 
Let us represent $\chi (z)$ as $\chi (z)  =  \pi^{-z/2} \Gamma (z/2)    e^{i \arg \zeta (z)}   |\zeta (z)|  $.    
 Define $\chitilde (z)$  as  $\chi (z)$ with 
the Stirling formula approximation to $\Gamma(z/2)$.     This approximation breaks the 
$z \to 1-z$ symmetry,  but since we know it exists,  we can easily  restore it: 
\beq
\label{restore} 
   \chi (z)  \approx    \smallhalf \[ \chitilde (z)  + \chitilde (1-z) \]
   \eeq
 One
    can show that for large $y$, 
\barray
\label{chiapprox}
  \chitilde (a + i y)  &=&   \sqrt{2}   \,  \pi^{(1-a)/2}  \(  y/2  \)^{(a-1)/2} e^{-y \pi/4} 
   |\zeta (a + i y)|  
\\  \nonumber 
&~& ~~~~\times   \exp\[ i \(  \frac{y}{2} \log \( \frac{y}{2 \pi e} \)  + \frac{ (a-1)\pi }{4} + \arg \zeta (a + i y) +O(1/y)  \) \]
\earray

Let us now show for  $a\neq \smallhalf$,   this approximation  $\chi (a + i y)
= \smallhalf ( \chitilde (a+iy)  + \chitilde (1-a - iy) )$  for large  $y$ 
has no zeros,  since its real and imaginary parts,   which just involve  sine's and cosine's of
the $\arg$-function   of the exponential in eq. (\ref{chiapprox}),  cannot simultaneously be zero.  
   To be more specific,  requiring that  both the real and imaginary parts
of   this approximation to $\chi (a + i y)$  be zero  leads to the two equations 
\beq
\label{sincos}
\cos \theta + b \cos \theta' =0,   ~~~~~ \sin \theta - b \sin \theta' =0
\eeq
  where $\theta$ 
is the argument of the exponential in eq.  (\ref{chiapprox}),  $\theta' = \theta (a \to 1-a)$,  
and $b = A(1-a)/A(a)$ where $A$ is the overall real factor multiplying the exponential in the above
equation.   In  order to be more rigorous,  one should shift away from a zero with a small complex 
number 
$\delta$,  i.e. consider  $\lim_{\delta \to 0} \chi (\rho +\delta)=0$.   In this way,  when $\delta$ is
not strictly zero,  one can first  safely divide by $A$ in order to obtain eq. (\ref{sincos}),  and
then take the limit $\delta \to 0$.     Squaring these two equations,  one finds that they 
 imply $\cos (\theta + \theta') = - (b +1/b) /2$.   However since $b+1/b \geq  2$  for 
any $b$,  there are no solutions,  except for $b=1$.     
Although as one approaches a zero,  both $A(1-a)$ and $A(a)$ are zero because of
the $|\zeta (z)|$ factor,  their ratio is well defined and indeed equals $1$,  so the above argument
is consistent.   

  The solution of the above equations (\ref{sincos}),  with $b=1$,  that corresponds to the 
zeros is 
\beq
\label{solu}
\theta = \theta'   , ~~~~~\cos \theta =0.  
\eeq
This is not the most general solution to (\ref{sincos}),  but the other solutions  are inconsistent with
the functional relation.    The constraint $\theta = \theta'$  implies $a= \smallhalf$.   
Thus for large $y$,  there are no Riemann zeros off of the critical line.  
 
 The above analysis leads to an explicit equation satisfied by the $n$-th zero which depends only
 on $n$.  
The equation (\ref{solu}) shows that the Riemann zeros are in one-to-one correspondence with
the zeros of the cosine,   i.e. occur when  
 $ \theta = (m- \smallhalf) \pi$,  where $m$ is an integer.     In order to conform with
the standard definition of the $n$-th zero by comparing with the lowest zeros,   one must
chose $m=n-1$,  and one finds
that the $n$-th  Riemann zero  is of the form $\rho_n = \smallhalf + i y_n$,  $n=1,2,3,...$, where 
$y_n$   satisfies the following equation: 
\beq
\label{FinalTranscendence} 
  \frac{y_n}{2 \pi}  \log \( \frac{y_n }{2 \pi e} \)     + \inv{\pi} \arg \zeta \( \smallhalf
+ i y_n \)   = n - \frac{11}{8}
\eeq
where we have dropped the $O(1/y_n)$ terms.   

It follows that   eq.  (\ref{FinalTranscendence})  implies that the  number of zeros along the critical line with $\Im (\rho) < T$ is given by
the above expression with $n = N+\smallhalf$ and $y_n =T$:
\beq
\label{counting2}
N(T)  =   \frac{T}{2 \pi}  \log \( \frac{T}{2 \pi e} \)   + \frac{7}{8}  + \inv{\pi} \arg \zeta \( \smallhalf
+ i T \)  + O(1/T)
\eeq
The result above  was  obtained by Riemann for the zeros on the entire 
critical strip\cite{Conrey}.    The $\arg \zeta$ correction is due to Backlund\cite{Edwards}.
Riemann  knew about the additional $7/8$.   The latter has also been obtained from quantization
of the Berry-Keating hamiltonian\cite{BerryKeating,German78,Bhaduri}.   
However,    eq. (\ref{FinalTranscendence}) is a stronger result since it is for zeros
on the critical line and is an equation satisfied by each  individual zero which depends only 
on $n$.      
It should be stressed that although the two equations   (\ref{FinalTranscendence})  and (\ref{counting2})  resemble each other,  it is impossible to derive the former from the latter,
again because the latter has only been proven for the entire strip assuming $T$ is not the ordinate of a zero.      It is clearly stated in Edward's book that $N(T)$ has never been shown to be valid on the critical line.

\def\smallelev{{\scriptstyle  \frac{11}{8}}}

The $\arg \zeta $  is small compared to the other terms.  
This leads to a simple approximation of the $n$-th zero on the critical line, 
Dropping this $\arg$ term, 
$y_n$ is  a solution of the transcendental equation 
\beq
\label{transcends}
n =  \frac{y_n}{2 \pi}   \log \( \frac{y_n}{2 \pi e} \)  + \frac{11}{8}
\eeq  
The solution is explicitly given in terms of Lambert's W-function\cite{Franca}
\beq
\label{Lambert}
y_n =    \frac{ 2 \pi \(n-  11/8 \)}{W \(  ( n - 11/8) /e  \) }  
\eeq
Although this just a simple inversion,  it is quite useful since the Lambert function is
implemented in most numerical packages.      The above formula neither misses zeros nor
predicts non-existent zeros,   i.e. every zero is in the vicinity of $y_n$ for some $n$,  
and there are no zeros between $n$ and $n+1$.   

The above formula is  closely related, but not identical,    to the  Gram points $g_n$,  which are solutions 
to $\vartheta (g_n ) = n \pi$  where $\vartheta$ is the Riemann-Siegel
 $\vartheta$ function.    We have not used the function $\vartheta$  in the analysis of this paper. 
  The Gram point that is closest to the first zero is  $g_0 = 17.8455$,
 compared to our $y_1 = 14.52$ which is much closer to the true zero.  
Gram's Law is  the tendency for Riemann zeros to   lie between consecutive Gram points,  but 
it is known to fail for about $1/4$ of all Gram intervals\cite{Edwards}.           
Our result is essentially different,  in that the eq.   (\ref{transcends})  is an   asymptotic formula  for 
the actual zeros and is thus  somewhat stronger than Gram's criterion,   
which in any case is  known to be
violated.

   Table \ref{Table1}  shows how well the formula (\ref{Lambert}) works 
for  various  $n$ up to $10^{22}+1$,  where for the latter the fractional  error is about $10^{-22}$.   This  confirms
the validity of the approximations we have made.   
Tables of  zeros were  calculated to high accuracy by Odlyzko,  which in part relies on the Gram points,    and  can be found at  his website \cite{Odlyzko}.  The two adjacent zeros at 1 million indicates that  eq. (\ref{transcends})  can accurately
distinguish nearby zeros.
Here is our estimate of the $10^{10^6}$-th zero:
\beq
\label{10tomillion}
\rho =  \smallhalf+ i\, 2.72877125379720787388146263022827376095518195769921562... \times 10^{999994}
\eeq
and is  exact to the number of digits shown,  and actually much more:   the Lambert function approximation is here accurate to a million digits,  which we were easily able to calculate with
Mathematica.      The approximation
(\ref{Lambert})  for the $10^p$'th zero is correct to roughly $p$ digits.    

\begin{table}
\begin{center}
\begin{tabular}{|c|c|c|}
\hline\hline
n &  ~ ~~exact $y_n$~~ ~ &~~~ solution to eq. (\ref{transcends}) ~  ~ ~~ \\
\hline\hline 
10    & 49.8  &  50.2  \\
100 & 236.5 &236.0 \\
1000    &  1419.4    &  1419.5   \\
10,000  &  9877.8     &9877.6  \\
100,000     &  74920.8   &  74920.9 \\
1,000,000    & 600269.7    & 600269.6  \\
1,000,001    &600270.3   & 600270.2 \\
10,000,000   &   4992381.0    &  4992381.1 \\
$10^{12}+1$  &  267653395648.8  & 267653395649.0\\
$10^{22} +1$  &           1370919909931995308226.68                   &1370919909931995308226.77\\
\hline\hline 
\end{tabular}
\end{center}
\caption{\emph{ The $n$-th Riemann zero  $\rho_n = \smallhalf + i \, y_n$ for various $n$ up to $10^{22}+1$.    The second column is exact to
the number of digits shown.      The third column is the explicit  formula  
(\ref{Lambert}) in terms of the Lambert function,   which is an approximation to 
the equation (\ref{FinalTranscendence}).
  The last  2 entries are 
from  Odlyzko's tables\cite{Odlyzko}.}  } 
\label{Table1}
\end{table}

The $\arg \zeta (\smallhalf + i T )$  term in  eq. (\ref{counting2})   oscillates around zero,
and changes sign in the vicinity of  each Riemann zero.  
It is thus responsible for turning the smooth part of $N(T)$ into a staircase function that
actually counts the zeros\cite{Berry1,BerryKeating}.    
 At a zero it can be defined by  the well-defined 
limit  $\lim_{\delta \to 0}  (\[ \arg \zeta (\smallhalf + i (y + \delta)) +  \arg \zeta (\smallhalf + i (y - \delta)) \] /2$.    The latter is generally not zero.    
 For instance, for the first Riemann zero, 
$\arg \zeta\(\tfrac{1}{2} + i y_1\) = 0.1578739\dotsc$.   
Equation (\ref{FinalTranscendence})    indeed   improves the estimates
of the Riemann zeros of the last section.    It is more difficult to solve numerically, 
but simple root-finder software such as in Mathematica can easily solve the equation 
 by searching for a solution in the vicinity of the approximation in 
eq. (\ref{Lambert}).  
The result 
is surprisingly accurate.      For the first zero it gives 
\beq
\label{firstzero}
y_1  =   14.13472514173469379045725198356247
\eeq
which is correct to the number of digits shown\cite{Odlyzko}. 
This strongly suggests that the $\CO(1/y_n)$ corrections in eq. (\ref{FinalTranscendence}) 
are completely under control.  

Montgomery's  pair correlation conjecture\cite{Montgomery} relates the statistics of the zeros 
to  that of random Hermitian
matrices in the GUE universality class\cite{Dyson}.   Strong evidence for this conjecture was 
given  by Odlyzko\cite{Odlyzko2}.   It is known that 
the  fluctuating $\arg \zeta$ correction in eq.   (\ref{FinalTranscendence})    is important  for such statistics\cite{Berry1,BerryKeating}.

Let us return to the statement that the validity of the RH is equivalent to the property that 
$f(y)$  of the last section is a regular alternating function.      Away from the pole at  $z=1$,  as in section II  
 we can define an 
electric field $\Evec'$  from the real and imaginary parts of $\chi (z)$,  which also has zero
divergence and curl,   and  $\Evec' = - \gradvec \phi'$.      The same arguments as in section 
II apply,  and the RH follows if $\Im (\chi (1 + i y))$  is a regular alternating function.    
For large $y$,   this   follows from the log-periodic behaviour in eq.  (\ref{chiapprox}),  
i.e. $\cos (y \log y)$.   

The formulas of this section can also be used to demonstrate the statement made in section II
that the vorticity of the zeros alternate in sign as one moves up the critical line.    
The latter implies that the electric field,  which is horizontal along the critical line,   
changes direction at a zero,  but between zeros is always either to the left or right,
as depicted in Figure \ref{Electric}.    This follows again from the $\log$ periodic 
behavior of $\chi$ along the critical line,   i.e.   $\chi (\smallhalf + iy ) \sim \cos \( 
y \log (y/2 \pi e)/2 \)$.

\section{Concluding Remarks}

In summary,  our analogy with the  electric field $\Evec$ and potential $\phi$  suggested a very regular pattern of
the function $\xi (z)$  in the critical strip.  We argued that the RH follows if the real function $\phi$ on
the line $\Re (z) =1$,  or
equivalently  $\Im (\xi (1+i y))$, is
what we referred to as a regular alternating function of $y$. 
     This motivated the analysis of the last   section,  which 
revealed such a pattern,  more specifically the  log-periodic behavior we derived in eq.  (\ref{chiapprox}).     
This  led to the simple formulas  (\ref{transcends}) and (\ref{FinalTranscendence})   for the n-th Riemann zero at large $n$.   The approximations in section IV that led to  eq. (\ref{FinalTranscendence}) 
are well controlled,  as indicated by the fact that accurate values for even the lowest zeros
can be obtained from it.     

  The  equations   
   (\ref{FinalTranscendence}) 
and   (\ref{transcends})  are  equations that determine the  zeros
on the critical line,  and Riemann's counting formula eq. (\ref{Nzeros}) 
 for the zeros on the entire strip  is a consequence of them,   
 including the corrections in
eq.  (\ref{counting2}).       
   We also provided additional analysis 
in the last  section showing that there are no zeros off of the critical line with  sufficiently large
imaginary part.   
Our work  thus indicates an extreme regularity of the zeros  with large imaginary part. 
Since it is already known that there are no zeros off the line for  $\Im (z) < 10^9$ or more\cite{Odlyzko,Conrey},
this would seem to establish the RH,  especially since the equation  (\ref{FinalTranscendence}) 
is nearly identically satisfied even for the lowest zeros.

\bigskip 

{\it Note added:}     We have recently published two follow-up's  to this article\cite{Gui}  
wherein we show that the solutions of eq. (\ref{FinalTranscendence}) are accurate enough
to reveal the GUE statistics.   We also provide a more rigorous treatment of section IV,
and derive an exact version of eq. (\ref{FinalTranscendence}),  which completely controls the
$1/y_n$ corrections.     We have also generalized the 
equations of the last section   to arbitrary Dirichlet L-functions\cite{Dirichlet}.

\bigskip

\section{Acknowledgments}  

I  especially wish  to thank Giuseppe Mussardo for  useful discussions 
 and the hospitality of the ICTP and SISSA in 
Trieste, Italy,  where this work was nearly    completed.    Germ\'an Sierra and Andrew Odlyzko  provided useful comments on the manuscript,  in particular the similarity of our $y_n$  with Gram points which I was 
 previously unaware of.   In particular,  Sierra   suggested  to try and obtain the fluctuating 
 part in $N(T)$ which we presented in section IV.      I also wish to thank Guilherme Fran\c ca 
 for subsequent  discussions  on the GUE statistics of the Riemann zeros.   
 
 I also wish to thank Olus Boratav,  for  challenging  me a few years ago to write a popular article on
the Riemann Hypothesis,  
 and this only   recently re-awakened my interest.     In any case,  there is already an excellent popular 
 book on the subject\cite{Derbyshire}.  
   This work is supported by the National Science Foundation of the United States of America 
under grant number  NSF-PHY-0757868.

\vfill\eject


\begin{thebibliography}{99}
\bibitem{sigma}
 In the mathematics literature
the complex variable $z$ is conventionally  denoted as $s= \sigma + i t$ in this context.  Here the notation will be $s=z=x + i y$ since this makes the vectorial aspects we will consider more natural.
\bibitem{Rpaper}     A translated version of his paper can be found in \cite{Edwards}.
\end{thebibliography}
\end{document}